\documentstyle[epsfig,12pt]{article}
\newlength{\dinwidth}
\newlength{\dinmargin}
\newcommand{\ba}{\begin{array}}
\newcommand{\ea}{\end{array}}
\newcommand{\be}{\begin{equation}}
\newcommand{\ee}{\end{equation}}
\newcommand{\bea}{\begin{eqnarray}}
\newcommand{\eea}{\end{eqnarray}}

\def\bra{\langle}
\def\ket{\rangle}

\def\a{\alpha}
\def\b{\beta}
\def\g{\gamma}

\def\p{\pi}

\def\l{\lambda}
\def\m{\mu}
\def\n{\nu}

\def\to{\rightarrow}
\begin{document}
\thispagestyle{empty}
\addtocounter{page}{-1}
\begin{flushright}
DESY 97-235\\
KAIST-TH 12/97\\
BUTP-97/34\\
SNUTP 97-169\\
December 1997
\end{flushright}
\vspace*{1.8cm}
\begin{center}
{\large\bf Contribution of $\mathbf b\rightarrow sgg$ through the QCD 
  anomaly\\
in exclusive decays $\mathbf B^{\pm}\rightarrow
  (\eta^{\prime},\eta)(K^{\pm},   K^{*\pm})$ and
$\mathbf B^{0}\rightarrow (\eta^{\prime},\eta)(K^{0},
  K^{*0})$  }
\end{center}
\vspace*{1.0cm}
\centerline{\large\bf A. Ali$^a$, J. Chay$^b$, C. Greub$^c$
 and  P. Ko$^d$}
\vspace*{0.5cm}
\centerline{\large\bf {\rm ${}^a$}Deutsches Elektronen Synchrotron
DESY,  Hamburg, Germany}
\centerline{\large \bf {\rm ${}^b$}Department of Physics, Korea
  University, Seoul 136-701, Korea}
\centerline{\large \bf {\rm ${}^c$} Inst. f. Theor. Physik, Univ.
Bern, Bern, Switzerland}
\centerline{\large \bf {\rm ${}^d$} Department of Physics, KAIST,
Taejon 305-701, Korea}
\vspace*{1.0cm}
\centerline{\Large\bf Abstract}
\vspace*{1cm}
We compute the decay rates for the exclusive decays $B^{\pm}
\rightarrow (\eta^{\prime},\eta) (K^{\pm}, K^{*\pm})$ and
$B^{0}\rightarrow (\eta^{\prime},\eta) (K^{0}, K^{*0})$ 
in a QCD-improved factorization
framework by including the contribution
from the process $b\rightarrow sgg \rightarrow s (\eta^{\prime},
\eta)$ through the QCD anomaly. This method provides an alternative
estimate of the contribution $b \to s c\bar{c} \to
s(\eta,\eta^\prime)$ to these  
decays as compared to the one using the intrinsic charm
content of the $\eta^{\prime}$ and $\eta$ mesons determined through
the decays $J/\psi \to (\eta,\eta^\prime ,\eta_c) \gamma$. The
advantage of computing the relevant matrix elements via the QCD
anomaly governing the transition $gg\rightarrow (\eta^{\prime},
\eta)$ is that there is no sign ambiguity in these contributions
relative to the matrix elements from the rest of the operators in the
weak effective Hamiltonian. Numerically, the QCD anomaly method and
the one using the radiative decays $J/\psi \to
(\eta,\eta^{\prime},\eta_c) \gamma$ give similar branching ratios for
the decays of interest here. The resulting branching ratios  are
compared with the CLEO data on $B^{\pm} \rightarrow \eta^{\prime}
K^{\pm}$ and $B^{0} \rightarrow \eta^{\prime} K^{0}$ and predictions
are made for the rest. 

\vspace*{1.5cm}
\centerline{(Submitted to Physics Letters B)}

\newpage

{\bf 1. Introduction}

\vspace{0.3cm}

 The CLEO collaboration has recently reported
measurements in a number
of exclusive two-body non-leptonic decays of the type $B\rightarrow
h_1 h_2$, where $h_1$ and $h_2$ are light mesons and the inclusive
decay $B^{\pm} \rightarrow \eta^{\prime}X_s$ \cite{CLEO1} -
\cite{CLEO3}. In particular,  large branching ratios into the final
states including $\eta^{\prime}$ are reported \cite{CLEO1,CLEO3}:
\begin{eqnarray}
\label{etapkpm}
B( B^{\pm} \rightarrow \eta^{\prime} + X_s ) &  = & (6.2 \pm 1.6 \pm
1.3 ) \times 10^{-4} ~({\rm for} ~2.0~{\rm GeV} \le p_{\eta^{'}}  \le
2.7~ {\rm GeV}),
\\
B( B^\pm \rightarrow \eta^{\prime} + K^\pm) & = & (6.5_{- 1.4}^{+ 1.5}
\pm  0.9) \times 10^{-5},
\\
B(B^0 \to \eta^{\prime} K^0 ) & = & (4.7_{-2.0}^{+2.7} \pm 0.9) \times 
10^{-5}~. 
\end{eqnarray}
Interestingly, no decay involving the $(\eta K)$ or
$(\eta,\eta^\prime)K^*$ modes of either the charged $B^\pm$ or the
neutral $B^0(\overline{B^0})$ has been observed and the corresponding
limits on some of these decays can be seen in \cite{CLEO3}. Of these,
the most  stringent limit is reported on the decay 
 $B^\pm \to \eta K^\pm$, for which, at $90\%$ C.L., one has 
\cite{CLEO3} 
\begin{equation}
\label{etapkzst}
B(B^\pm \to \eta K^\pm ) \leq 1.4 \times 10^{-5} .
\end{equation}
These measurements have stimulated a lot of theoretical activity 
\cite{AG97} - \cite{Kim97}.
 We will concentrate in this paper on
the exclusive two-body decays $B \to (\eta^{\prime},\eta)(K,K^*)$ for
both neutral and charged $B$ mesons.

In ref..~\cite{AG97},  two of us (A.A. and C.G.) have studied a number 
of  non-leptonic two-body exclusive decay modes of the $B^\pm$ and
$B^0$ mesons  in the QCD-improved effective Hamiltonian approach,
involving the effective four-quark and magnetic moment operators.
 Since the relevant matrix elements
of the type $\langle h_1 h_2| {\cal O}_i | B\rangle$, where
${\cal O}_i$ are four-quark operators, are difficult to
estimate from first principles, one often resorts to
the factorization approximation \cite{BSW87}, in which the matrix
elements of interest factorize into a product of two relatively more
tractable hadronic matrix elements. The resulting decay rates depend
on a set of effective parameters, which have to be determined from
experiments. We recall that this generalized factorization approach
appears to describe the two-body non-leptonic $B$ decays involving the
so-called heavy to heavy transitions  reasonably well
\cite{NS97}. Likewise, data on $B \to K \pi$ and $B \to \pi \pi$
decays are well accounted for in this framework \cite{AG97}.

In contrast to the decays $B \to K\pi$ and $B \to \pi \pi$, the decays
 $B \to (\eta,\eta^{\prime})(K,K^*)$
in the factorization approach require additionally
the knowledge of the matrix elements $\langle \eta^{\prime}
|\overline{c}\gamma_{\mu} \gamma_5 c|0\rangle$ and
$\langle \eta |\overline{c}\gamma_{\mu} \gamma_5 c|0\rangle $,
emerging from the decay $b \to s (\bar{c} c) \to s
 (\eta^{\prime},\eta)$. Parameterizing them as $\langle \eta^{\prime} 
|\overline{c}\gamma_{\mu} \gamma_5 c|0\rangle =
-if_{\eta^{\prime}}^{c} q_{\mu}$ and $\langle \eta
|\overline{c}\gamma_{\mu} \gamma_5 c|0\rangle =
-if_{\eta}^{c} q_{\mu}$, the quantities of interest
for this contribution are $f_{\eta^{\prime}}^{c}$ and
$f_{\eta}^{c}$, which are often referred to as the charm content of
 the  $\eta^{\prime}$ and $\eta$, respectively \cite{Berkelman}.
 These quantities are {\it a priori} unknown but they can
be determined in a number of ways, also including the $B$-decays 
being discussed here. In \cite{AG97}, these quantities were determined
from the decays $J/\psi  \to (\eta,\eta^{\prime},\eta_c) \gamma$,
extending the usual $(\eta,\eta^{\prime})$-mixing formalism
\cite{GK87} to the $(\eta_c,\eta^{\prime},\eta)$ system. Using the
measured  decay widths for the decays
$J/\psi \to (\eta,\eta^{\prime},\eta_c) \gamma$ 
and $(\eta_c,\eta^{\prime},\eta) \to \gamma \gamma$ yields
$|f_{\eta^\prime}^c| \simeq 5.8$ MeV
and $|f_{\eta}^c| \simeq 2.3$ MeV  \cite{AG97}.

 In the meanwhile,
theoretical arguments based on $SU(3)$-breaking effects in the 
pseudoscalar nonet
$(\pi,K,\eta,\eta^{\prime})$  in the chiral
perturbation theory approach \cite{Leutwyler97}, and phenomenological
analysis involving
in particular the $\eta\gamma$ and $\eta^{\prime}\gamma$ transition
form  factors \cite{FK97},
have put to question the conventional (one mixing-angle) octet-singlet 
mixing scheme for
the $(\eta,\eta^\prime)$ system. The modified two-angle mixing scheme,
proposed in \cite{Leutwyler97}, has also
implications for $B$ decays involving the $\eta$ and $\eta^\prime$
meson in the final state.
In particular, estimates of $|f_{\eta^\prime}^c|$ and
$|f_{\eta}^c|$ are expected to get revised.
Numerically, these quantities depend on the input values of
the pseudoscalar coupling constants, $f_0$ and $f_8$, and the two
mixing angles, called $\theta_0$ and $\theta_8$. However, it is found
that using the best-fit values of the parameters from  \cite{FK97},
which are consistent with Leutwyler's estimates of the same
\cite{Leutwyler97}, the estimate of $|f_{\eta^\prime}^c|$ in the modified
mixing scheme remains practically unaltered. In contrast, the quantity
$|f_{\eta}^c|$ is considerably reduced due to the small value of the
mixing angle in the singlet sector,
which makes $\eta$  an almost octet state. One finds now
$|f_{\eta}^c| \simeq 0.9$ MeV \cite{AG97}.
 With these estimates, it has been argued in \cite{AG97}  that
the charm-induced contribution $b \to s(c\bar{c}) \to
s(\eta^\prime,\eta)$  does not dominate the matrix element for
$B^{\pm} \rightarrow \eta^{\prime}  K^{\pm}$. The resulting branching
ratio $BR(B^\pm \to \eta^\prime K^\pm)=  (2 -4)\times 10^{-5}$ is
somewhat lower than but not inconsistent with the experimental number
in Eq.~(2).  

 The  branching ratio for $B^{\pm} \rightarrow \eta^{\prime}
K^{\pm}$
(and other related decay modes) depends on the sign of the quantity 
$f_{\eta^\prime}^c$
(and $f_{\eta}^c$ involving the $\eta$ meson), as well as on the 
phenomenological parameter $\xi$,
which in turn determines the effective Wilson coefficients 
in the factorization approach \cite{AG97}.
It is not unreasonable to expect that the value of $\xi$ will be
similar in the decays $B \to h_1 h_2$, as the energy released in these
decays are comparable; hence this parameter can be determined in a
number of $B$ decays  \cite{AKL97}. A
consistent determination of the parameter $\xi$ will also check the
consistency of the underlying theoretical framework, namely
QCD-improved factorization.  However, it is desirable to
get independent estimates of the quantities $f_{\eta^\prime}^c$ and 
$f_{\eta}^c$, and
also settle the sign ambiguity present in the method used in
\cite{AG97}. We note that a recent phenomenological study has put a
bound on $f_{\eta^\prime}^c$, namely $-65 ~\mbox{MeV} \leq
f_{\eta^\prime}^c \leq 15  ~\mbox{MeV}$,
with $f_{\eta}^c$ being consistent with zero \cite{FK97} by analyzing
the $Q^2$ evolution of the $\eta\gamma$ and $\eta^\prime \gamma$ form
factors, respectively.
The bounds on $f_{\eta^\prime}^c$ from this method are
not very  stringent and the estimates of this quantity in \cite{AG97}  
are well within these bounds.

 In this letter we propose another
 method for computing the contribution of the amplitudes
$b\rightarrow s (gg) \rightarrow s(\eta^{\prime},
\eta)$. This method is based on calculating the amplitude for the
chromomagnetic penguin process
$b\rightarrow s gg$, followed by the transitions $gg\rightarrow
(\eta^{\prime}, \eta)$ which are calculated using the
QCD anomaly, determining both the sign and magnitude of
these contributions. As discussed below, the numerical values of 
$f_{\eta^\prime}^c$ and $f_{\eta}^c$ now depend on the charm quark
 mass (both of them being essentially proportional to $m_c^{-2}$).
Varying $m_c$ in the range $1.3$ - $1.5$ GeV, we
find that the QCD-anomaly-method gives
$f_{\eta^\prime}^c = -3.1$ ($-2.3$) MeV and $f_{\eta}^c = -1.2$ ( $
- 0.9$) MeV. Hence,
in absolute value, $f_{\eta}^c$ turns out to be very close to the one 
obtained in
the $(\eta_c,\eta^\prime,\eta)$-mixing formalism \cite{AG97} (the two 
estimates
almost coincide for $m_c =1.5$ GeV), but the value of
$f_{\eta^\prime}^c$ is typically a factor 2 smaller in the
QCD-anomaly method.  
The branching ratios for $B^\pm \to (\eta^\prime,\eta)(K^\pm,K^{*\pm})$
and $B^0 \to (\eta^\prime,\eta)(K^0,K^{*0})$ 
based on the QCD-anomaly method are calculated in this letter and 
compared with the present CLEO measurements and with the ones in
 \cite{AG97}. We find that the theoretical branching ratios for $B^\pm
 \to \eta^\prime  K^\pm$ and
$B^0 \to \eta^\prime K^0$ are almost equal and both are in the range
 $(2 -4)  \times 10^{-5}$, in agreement with the estimates in
 \cite{AG97}.  

\vspace{0.3cm}

{\bf 2. Estimate of ${\mathbf b \to (\eta,\eta^\prime)s}$ via QCD
anomaly} 

\vspace{0.3cm}
We write the  effective Hamiltonian $H_{\mathrm{eff}}$ for the
$\Delta B=1$ hadronic transitions as
\be
\label{heff}
H_{\mathrm{eff}}
= \frac{G_{F}}{\sqrt{2}} \, \left[ V_{ub} V_{uq}^*
\left (C_1 O_1^u + C_2 O_2^u \right)
+ V_{cb} V_{cq}^*
\left (C_1 O_1^c + C_2 O_2^c \right) -
V_{tb} V_{tq}^* \, \left(
\sum_{i=3}^{10}
C_{i} \, O_i + C_g O_g \right) \right]  \quad ,
\ee
where
$q=d,s$. The operators read
\bea
\label{operators}
O_1^u &=& \left( \bar{u}_\a b_\a \right)_{V-A} \,
        \left( \bar{q}_\b u_\b \right)_{V-A} \quad \quad
O_1^c = \left( \bar{c}_\a b_\a \right)_{V-A} \,
        \left( \bar{q}_\b c_\b \right)_{V-A} \nonumber \\
O_2^u &=& \left( \bar{u}_\b b_\a \right)_{V-A} \,
        \left( \bar{q}_\a u_\b \right)_{V-A} \quad \quad
O_2^c = \left( \bar{c}_\b b_\a \right)_{V-A} \,
        \left( \bar{q}_\a c_\b \right)_{V-A} \nonumber \\
O_3 &=& \left( \bar{q}_\a b_\a \right)_{V-A} \, \sum_{q'}
        \left( \bar{q}'_\b  q'_\b \right)_{V-A} \quad
O_4  =  \left( \bar{q}_\b b_\a \right)_{V-A} \, \sum_{q'}
        \left( \bar{q}'_\a  q'_\b \right)_{V-A} \quad \nonumber \\
O_5 &=& \left( \bar{q}_\a b_\a \right)_{V-A} \, \sum_{q'}
        \left( \bar{q}'_\b  q'_\b \right)_{V+A} \quad
O_6  =  \left( \bar{q}_\b b_\a \right)_{V-A} \, \sum_{q'}
        \left( \bar{q}'_\a  q'_\b \right)_{V+A} \nonumber \\
O_7 &=& \frac{3}{2}
\left( \bar{q}_\a b_\a \right)_{V-A} \, \sum_{q'} e_{q'}
        \left( \bar{q}'_\b  q'_\b \right)_{V+A} \quad
O_8  =  \frac{3}{2}
\left( \bar{q}_\b b_\a \right)_{V-A} \, \sum_{q'} e_{q'}
        \left( \bar{q}'_\a  q'_\b \right)_{V+A} \quad \nonumber \\
O_9 &=& \frac{3}{2}
\left( \bar{q}_\a b_\a \right)_{V-A} \, \sum_{q'} e_{q'}
        \left( \bar{q}'_\b  q'_\b \right)_{V-A} \quad
O_{10}  =  \frac{3}{2}
\left( \bar{q}_\b b_\a \right)_{V-A} \, \sum_{q'} e_{q'}
        \left( \bar{q}'_\a  q'_\b \right)_{V-A} \quad \nonumber \\
O_g &=& (g_s/8\p^{2}) \, m_b \, \bar{s}_{\a} \, \sigma^{\m \n}
      \, (1+\g_5)  \, (\l^A_{\a \b}/2) \,b_{\b}
      \ G^A_{\m \n} \quad .
\eea
Here $\alpha$ and $\beta$ are the $SU(3)$ color indices and $\l^A_{\a
        \b}, A=1,...,8$, are the Gell-Mann matrices. The subscripts $V
        \pm A$ represent 
the chiral projections $1 \pm \gamma_5$. $G^A_{\m \n}$
denotes the gluonic field strength tensor. The operators
$O_1$ and $O_2$ are the current-current operators, $O_3,...,O_6$ the
so-called gluonic penguin operators and $O_7,...,O_{10}$ are
the electroweak penguin operators. Finally, $O_g$ is the gluonic
dipole operator. In the following we use the next-to-leading
logarithmic values (NLL) (with respect to QCD) for $C_1,...,C_6$,
computed in \cite{Burasmartinelli}, while
the remaining coefficients are taken at leading logarithmic precision
(LL). Taking $\alpha_s(m_Z)=0.118$, $\alpha_{ew}(m_Z)=1/128$
and $m_{top}^{pole}=175$ GeV,  the coefficients in the naive
dimensional renormalization scheme (NDR) (evaluated at the
renormalization scale $\mu=2.5$ GeV) read:
 $C_1=1.117$, $C_2=-0.257$,
$C_3=0.017$, $C_4=-0.044$, $C_5=0.011$, $C_6=-0.056$, $C_7=-1 \times
        10^{-5}$, 
$C_8= 5 \times 10^{-4}$, $C_9=-0.010$, $C_{10}=0.002$ and
$C_g=-0.158$.
Among the coefficients of the electroweak penguins only
$C_9$ has a sizable coefficient, which arises mainly from the
$Z^0$ penguin.

Working consistently to the precision mentioned above, we
include one-loop QCD corrections to the partonic matrix elements
of the operators $O_1,...,O_6$ and the tree-level diagram associated
with $O_g$, where the gluon splits into a quark-antiquark pair.
These  issues are discussed in detail in ref. \cite{AG97}.
In addition, we include the photonic penguin diagram associated with
the current-current operators $O_1$ and $O_2$.
 All these corrections
can be absorbed into effective Wilson coefficients
$C_i^{\mathrm{eff}}$ $(i=1,...,10)$
multiplying the
four-Fermi operators $O_1,...,O_{10}$ in the basis (\ref{operators}).
While $C_{1}^{\mathrm{eff}},...,C_{6}^{\mathrm{eff}}$ are given in
Eq.~(2.5) in ref. \cite{AG97}, the remaining effective coefficients
read  
\be
C_7^{\mathrm{eff}} = C_7 +\frac{\alpha_{ew}}{8\pi} C_e \quad ,
C_8^{\mathrm{eff}} = C_8 \quad ,
C_9^{\mathrm{eff}} = C_9 +\frac{\alpha_{ew}}{8\pi} C_e \quad ,
C_{10}^{\mathrm{eff}} = C_{10}  \quad ,
\ee
where
\be
C_e = -\frac{8}{9} (3C_2 + C_1) \, \sum_{q'=u,c} \,
\frac{V_{q'b}V_{q'q}^*}{V_{tb} V_{tq}^*} \left(
\frac{2}{3} + \frac{2}{3} \log \frac{m_{q'}^2}{\mu^2} -
\Delta F_1 \left( \frac{k^2}{m_{q'}^2} \right) \right) \quad .
\ee
The function $\Delta F_1$ is also given in ref. \cite{AG97}.
In the following, we often use the following linear combinations
of effective Wilson coefficients ($N_c=3$ is the number of colors):
\be
\label{adef}
a_i = C_i^{\mathrm{eff}} + \frac{1}{N_c} C_{i+1}^{\mathrm{eff}} \quad
(i=\mbox{odd})\quad; 
~a_i = C_i^{\mathrm{eff}} + \frac{1}{N_c} C_{i-1}^{\mathrm{eff}} \quad
(i=\mbox{even})\quad. 
\ee

As described in detail in ref.~\cite{AG97}, the hadronic
matrix elements $\bra h_1 h_2|C_i^{\mathrm{eff}} O_i|B\ket $ for the
two-body decays of the form $B \rightarrow h_1 h_2$
are now readily decomposed into various form factors
and decay constants when applying  the
factorization approximation. If $\eta^{(')}$ is involved in the
final state, the factorization of $O_1^c$ and $O_2^c$
brings in the matrix elements
\be
\bra \eta^{(')}|\bar{c} \gamma_\mu \gamma_5 c|0 \ket
\ee
which have to be estimated. We model them by annihilating
the charm-anticharm pair into two gluons, followed by the transition
$gg \rightarrow \eta^{(')}$ (see Fig.~1). The first part of
this two-step process, i.e. $b \rightarrow s
(\bar{c} c \rightarrow g(k_1) g(k_2))$
which amounts to calculating the charm-quark-loop from which two
gluons are emitted, has been worked out by Simma and Wyler
\cite{simma} in the context
of a calculation in the full theory. Their result is readily
translated to our effective theory approach and can be
compactly written as a new (induced) effective Hamiltonian
$H_{\mathrm{eff}}^{gg}$, 
\begin{equation}
  \label{newop}
H_{\mathrm{eff}}^{gg}=- \frac{\alpha_s}{2\pi} \left(
C_2^{\mathrm{eff}} + \frac{C_1^{\mathrm{eff}}}{N_c} \right)
\frac{G_F}{\sqrt{2}} V_{cb} V_{cs}^* \Delta i_5
\bigl(\frac{q^2}{m_c^2}\bigr)
\frac{1}{k_1 \cdot k_2} G^{\alpha \beta}_a (D_{\beta}
\tilde{G}_{\alpha \mu} )_a \, \overline{s}\gamma^{\mu} (1-\gamma_5) b
\quad , 
\end{equation}
with $\tilde{G}_{\mu \nu}=\frac{1}{2} \epsilon_{\mu \nu \a \b} G^{\a
\b}$ $(\epsilon_{0123}=+1)$.
In this formula, which holds for on-shell gluons
($q^2 = (k_1 + k_2)^2 = 2k_1 \cdot k_2$),
the sum over color indices is understood. The function $\Delta
i_5 (q^2/m_c^2)$ is defined as 
\begin{equation}
  \label{deltai5}
  \Delta i_5 (z) = -1 +\frac{1}{z} \Bigl[ \pi -2 \arctan (\frac{4}{z}
  -1)^{1/2} \Bigr]^2, \ \mbox{for $0<z<4$} \quad .
\end{equation}
Note that the $b \to s gg$ calculation brings in an explicit factor of
$\alpha_s$.
 However, as shown below, this explicit $\alpha_s$ 
factor gets absorbed into the matrix element of the operator resulting
from  the anomaly. So, to the order that 
we are working, we use the coefficients $C_1^{\mathrm{eff}}$ and
$C_2^{\mathrm{eff}}$ in 
eq. (\ref{newop}).
 By expanding the function  $\Delta i_5 (q^2/m_c^2)$ in inverse
powers of $m_c^2$, it is easy to see that the leading $(1/m_c^2)$ term
in Eq.~(\ref{newop}) generates the chromomagnetic analogue of the
operator considered by Voloshin \cite{voloshin97} to
calculate the power $(1/m_c^2)$ correction in the radiative
decay $B \to X_s \gamma$.
It should be remarked that the corresponding $u\bar{u}$ contribution in
Fig. 1 is suppressed due to the unfavourable CKM factors. The 
$t\bar{t}$ contribution is included in the effective Hamiltonian via 
the $bsgg$ piece present in the operator $O_g$. However, in the
factorization framework, the $bsgg$ term in $O_g$ does not contribute
to the decays discussed. So, the $c\bar{c}$ contribution in Fig. 1 is
the only one that survives.

The (factorizable) contribution from
Eq.~(\ref{newop}) in the decays $B \to (\eta,\eta^\prime)(K,K^*)$ is
of the same order (in $\alpha_s)$ as those of the other operators
taken into account in \cite{AG97}.
 It should be remarked that in calculating the amplitude $b \to s gg$,
there are more contributions in this order in $\alpha_s$ than what is
shown in Eq.~(\ref{newop}) and Fig.~1.
However, the diagrams where one of
the gluons is present in the final state yielding $b \to s g
(\eta,\eta^\prime)$ are not expected  to contribute significantly
to the exclusive two-body decays, but rather induce multibody
decays. Certainly,
these configurations have to be included in inclusive decays $B \to
(\eta,\eta^\prime) X_s$ \cite{AS97} but can be neglected in the
exclusive decays $B \to (\eta,\eta^\prime)(K,K^*)$.
 Likewise, configurations in which one of the gluons
emanates from the effective $bsg$ vertex and the other is
bremsstrahled  from the $b$ or $s$-quark (or from the spectator
anti-quark in the $B$ meson) to form an $\eta^\prime$ or $\eta$ are
non-factorizing contributions and they are ignored as 
the rest of the amplitudes are also calculated in the factorization 
approximation.

Working out the hadronic matrix element of Eq.~(\ref{newop})
using factorization, we now need to evaluate the matrix
elements
\be
\label{opnew}
\langle
\eta^{(\prime)} |  G_a^{\alpha \beta} (D_{\beta} \tilde{G}_{\alpha \mu, 
a}) |0\rangle \quad .
\ee
The operator in Eq.~(\ref{opnew}) can be written as
\begin{equation}
  \label{parder}
 G^{\alpha \beta}_a (D_{\beta} \tilde{G}_{\alpha\mu} )_a =
 \partial_{\beta} (G^{\alpha \beta}_a \tilde{G}_{\alpha \mu,a}) -
 (D_{\beta} G^{\alpha \beta})_a \tilde{G}_{\alpha \mu,a} \quad .
\end{equation}
We can discard the second term since it is suppressed
by an additional power of
$g_s$ which follows on  using the equation of motion, and
furthermore, the first
term is enhanced by $N_c$ in the large $N_c$ limit. The matrix elements
of $\partial_{\beta} (G^{\alpha \beta}_a \tilde{G}_{\alpha \mu,a})$ are
related to those of $G\tilde{G}$; more explicitly
\be
\partial_\beta \bra \eta^{(')}|G^{\alpha \beta}_a \tilde{G}_{\alpha
\mu,a} | 0 \ket = \frac{i q_\mu}{4}
\bra \eta^{(')}|G^{\alpha \beta}_a \tilde{G}_{\alpha \beta,a}|0 \ket
\quad .
\ee
The conversion of the gluons into $\eta$
and $\eta^{\prime}$ is described by an amplitude which is fixed by the
$SU(3)$ symmetry and the axial $U(1)$ current triangle anomaly. The
matrix elements for $G\tilde{G}$ can be written as \cite{voloshin80}
\be
\label{anomaly}
\langle \eta^{(')} | \frac{\alpha_s}{4\pi} G^{\alpha \beta}_a
\tilde{G}_{\alpha \beta,a} |0 \rangle = m_{\eta^{(')}}^2
f_{\eta^{(')}}^u \quad .
\ee
In Eqs.~(\ref{anomaly}) the decay constants $f_{\eta^\prime}^u$ and
$f_{\eta}^u$ read
\begin{equation}
\label{koppetas}
f_{\eta}^u = \frac{f_8}{\sqrt{6}} \, \cos \theta_8 -
               \frac{f_0}{\sqrt{3}} \, \sin \theta_0  \quad , \quad
f_{\eta^\prime}^u = \frac{f_8}{\sqrt{6}} \, \sin \theta_8 +
               \frac{f_0}{\sqrt{3}} \, \cos \theta_0  \quad , \quad
\end{equation}
where the coupling constants $f_8$, $f_0$ 
and the mixing angles $\theta_8$ and $\theta_0$
have been introduced earlier. We follow here the  two-angle
$(\eta,\eta^\prime)$ mixing formalism of ref.~\cite{Leutwyler97},
where the mass eigenstates $|\eta\rangle$ and $|\eta^\prime \rangle$
have the following decompositions: 
\begin{eqnarray}
\label{mix1}
|\eta \rangle = \cos \theta_8 |\eta_8 \rangle - \sin \theta_0 |\eta_0
\rangle , 
\nonumber\\ 
|\eta^\prime \rangle = \sin \theta_8 |\eta_8 \rangle + \cos \theta_0
|\eta_0  
\rangle . \end{eqnarray}

Collecting the individual steps,
the matrix elements in Eqs.~(\ref{opnew}) can be written as
\be
\label{opnewa}
\langle
\eta^{(\prime)}(q) |  \frac{\alpha_s}{4\pi}
G_a^{\alpha \beta} (D_{\beta} \tilde{G}_{\alpha \mu
a}) |0\rangle = i q_\mu \frac{m_{\eta^{(')}}^2}{4}
f_{\eta^{(')}}^u  \quad .
\ee

One would have naively expected that the gluonic matrix elements are
small since they contain an extra factor of $\alpha_s$. However, as
shown by  Eqs.~(\ref{opnewa}), this is obviously not the case and the
gluon operator with $\alpha_s$ as a whole is responsible for the
invariant mass of the  $\eta^{(\prime)}$ mesons. Also, the combination
entering in Eqs.~(\ref{opnewa}) involving the product of
$\alpha_s$ and the gluon field operators is independent of the
renormalization scale.

\vspace{0.3cm}
{\bf 3. Matrix elements for the decays ${\mathbf B^\pm \to
(\eta^\prime,\eta)(K^\pm,K^{*\pm})}$ and ${\mathbf B^0 \to
(\eta^\prime,\eta)(K,K^{*0})}$ }

\vspace{0.3cm}
 To compute the complete
amplitude for the exclusive decays, one has to
combine the contribution from the decay $b \to s (c\bar{c}) \to s(gg) \to 
s \eta^{(')}$ discussed in 
the previous section with all the others arising from the
four-quark and chromomagnetic operators, as detailed in \cite{AG97}.
The resulting amplitudes in the factorization approximation are listed
 below for all the eight cases of interest
$B^\pm \to (\eta^\prime,\eta)(K^\pm,K^{*\pm})$ and $B^0 \to 
(\eta^\prime,\eta)(K^0,K^{*0})$.
The expressions are given for the decays of the $B^-$ and
 $\overline{B^0}$ mesons; the ones for the charge conjugate decays are
 obtained by complex conjugating the CKM factors.

\newpage

${\mathbf (i)~B^- \to K^- \eta^{(')}}$

\begin{eqnarray}
\label{proc6}
M &=& \frac{G_F}{\sqrt{2}} \, \left\{ V_{ub} V_{us}^* \, \left[
a_2 + a_1 \frac{m_B^2-m_{\eta^{(')}}^2}{m_B^2-m_K^2} \,
\frac{F_0^{B \to \eta^{(')}}(m_K^2)}{F_0^{B \to
K^-}(m_{\eta^{(')}}^2)} \, 
\frac{f_K}{f_{\eta^{(')}}^u} \right]
-V_{cb}V_{cs}^* \, a_2 \Delta i_5
\Bigl(\frac{m_{\eta^{(')}}^2}{m_c^2}\Bigr)
\right. \nonumber \\
&& - V_{tb} V_{ts}^* \, \Biggl[2 (a_3 - a_5) + \frac{1}{2}(a_9-a_7) +
\Biggl( a_3 -a_5 -\frac{1}{2}(a_9-a_7) + a_4 -\frac{1}{2} a_{10}
 \nonumber \\
&& 
+ (a_6-\frac{1}{2}a_8) \frac{  m_{\eta^{(')}}^2}{ m_s(m_b-m_s)}
\Biggr) \, \frac{f_{\eta^{(')}}^s}{f_{\eta^{(')}}^u}
- (a_6 - \frac{1}{2} a_8) \frac{ m_{\eta^{(')}}^2}{ m_s(m_b-m_s)}
\nonumber \\ 
&& \left.  +
\left( a_4 +a_{10} + \frac{2 (a_6+a_8) m_K^2}{(m_s+m_u) \, (m_b-m_u)}
\right) \, \frac{m_B^2-m_{\eta^{(')}}^2}{m_B^2-m_K^2} \, \frac{F^{B
\to \eta^{(')}}_0(m_K^2)}{F^{B \to K^-}_0(m_{\eta^{(')}}^2)} \,
\frac{f_K}{f_{\eta^{(')}}^u} \Biggr] \,
\right\} \nonumber \\
&\times& \langle K^- | \bar{s} \, b_-|B^- \rangle \,
            \langle \eta^{(')} | \bar{u} \, u_-|0 \rangle
\end{eqnarray}
where
\begin{equation}
\label{proc6a}
 \langle K^- | \bar{s} \, b_-|B^- \rangle \,
            \langle \eta^{(')} | \bar{u} \, u_-|0 \rangle =
i \, f_{\eta^{(')}}^u \, (m_B^2 - m_K^2) \,
F_0^{B \to K^-}(m_{\eta^{(')}}^2).
\end{equation}
The coefficients $a_i$ are defined in Eq. (\ref{adef}) and the short-hand 
notation $ \bar{u} \, u_-$ stands for  $\bar{u} \, u_-=\bar{u} \gamma_\mu 
(1- \gamma_5)u$.
 The quantities $f_\eta^u$ and $f_{\eta^\prime}^u$ are 
given in Eqs. (\ref{koppetas}), while
$f_\eta^s$ and $f_{\eta^\prime}^s$ read
\begin{equation}
  \label{fetas}
f_{\eta^\prime}^s = -2 \, \frac{f_8}{\sqrt{6}} \, \sin \theta_8 +
               \frac{f_0}{\sqrt{3}} \, \cos \theta_0 \quad , \quad
f_{\eta}^s = -2 \, \frac{f_8}{\sqrt{6}} \, \cos \theta_8 -
               \frac{f_0}{\sqrt{3}} \, \sin \theta_0 \quad .
\end{equation}
Note that the matrix elements of the pseudoscalar density have been
expressed as
\begin{equation}
\label{density2}
\langle \eta^{(')} | \bar{s} \gamma_5 s | 0 \rangle = i
\frac{m_{\eta^{(')}}^2}{2 m_s}
(f_{\eta^{(')}}^u - f_{\eta^{(')}}^s) .
\end{equation}

\vspace*{1cm}

${\mathbf (ii)~B^- \to K^{*-} \eta^{(')}}$

\begin{eqnarray}
\label{proc8}
M &=& \frac{G_F}{\sqrt{2}} \, \left\{ V_{ub} V_{us}^* \, \left[
a_2 + a_1 \frac{F_1^{B \to \eta^{(')}}(m_{K^*}^2)}{A_0^{B
\to K^*}(m_{\eta^{(')}}^2)} \,
\frac{f_{K^*}}{f_{\eta^{(')}}^u} \right]
- V_{cb}V_{cs}^* \, a_2 \, \Delta i_5 \Bigl(
\frac{m_{\eta^{(')}}^2}{m_c^2} \Bigr)
\right. \nonumber \\
&& - V_{tb} V_{ts}^* \, \Biggl[2( a_3 - a_5)+
\frac{1}{2}(a_9-a_7) + \Biggl( a_3 -a_5- \frac{1}{2}(a_9-a_7) + a_4
-\frac{1}{2} a_{10} \nonumber \\
&& 
- (a_6 -\frac{1}{2}a_8) \frac{ m_{\eta^{(')}}^2}{ m_s(m_b+m_s)}
\Biggr) \, 
\frac{f_{\eta^{(')}}^s}{f_{\eta^{(')}}^u}
+ (a_6 -\frac{1}{2}a_8) \frac{  m_{\eta^{(')}}^2}{ m_s(m_b+m_s)}
\nonumber \\ 
&& \left.  +
(a_4 +a_{10})
\,  \frac{F^{B \to
\eta^{(')}}_1(m_{K^*}^2)}{A^{B \to K^*}_0(m_{\eta^{(')}}^2)}
\, \frac{f_{K^*}}{f_{\eta^{(')}}^u} \Biggr] \,
\right\} \, \langle K^{*-} | \bar{s} \, b_-|B^- \rangle \,
            \langle \eta^{(')} | \bar{u} \, u_-|0 \rangle \nonumber \\
\end{eqnarray}
with
\begin{equation}
\label{proc8a}
\langle K^{*-} | \bar{s} \, b_-|B^- \rangle \,
\langle \eta^{(')} | \bar{u} \, u_-|0 \rangle = -i \, f_{\eta^{(')}}^u
\, 2 m_{K^*} \, (p_B \epsilon_{K^*}^*) \, A_0^{B \to
K^*}(m_{\eta^{(')}}^2). 
\end{equation}

${\mathbf (iii)~\overline{B^0} \to \overline{K^0} \eta^{(')}}$

\begin{eqnarray}
M&=& \frac{G_F}{\sqrt{2}} \Biggl\{ V_{ub} V_{us}^* a_2 -V_{cb}
V_{cs}^* a_2 \Delta i_5 \Bigl( \frac{m_{\eta^{(')}}^2}{m_c^2} \Bigr) 
\nonumber \\
&-&V_{tb} V_{ts}^* \Biggl[ 2(a_3 -a_5) +\frac{1}{2}(a_9-a_7)+
\Biggl( a_3 -a_5 -\frac{1}{2}(a_9-a_7)+ a_4 -\frac{1}{2}a_{10}
\nonumber \\
&& +
(a_6 - \frac{1}{2} a_8)\frac{m_{\eta^{(')}}^2}{m_s (m_b - m_s)}
\Biggr) \frac{f_{\eta^{(')}}^s}{f_{\eta^{(')}}^u}
-(a_6-\frac{1}{2}a_8)\frac{ m_{\eta^{(')}}^2}{m_s (m_b - m_s)}
\nonumber \\ 
&+& \Bigl( a_4 -\frac{1}{2} a_{10} +
\frac{(2a_6-a_8) m_K^2}{(m_s + m_d)(m_b -m_d)} \Bigr)
\frac{m_B^2 -m_{\eta^{(')}}^2}{m_B^2 - m_K^2}
\frac{F_0^{B\rightarrow \eta^{(')}} (m_K^2)}{F_0^{B\rightarrow K}
(m_{\eta^{(')}}^2)} \frac{f_{K_0}}{f_{\eta^{(')}}^u} \Biggr]
\Biggr\} \nonumber \\
&\times& \langle \overline{K^0} |\overline{s} b_- |\overline{B^0}
\rangle \langle \eta^{(')} |\overline{u} u_-|0\rangle.
\end{eqnarray}

${\mathbf (iv)~\overline{B^0} \to \overline{K^{*0}} \eta^{(')}}$

\begin{eqnarray}
M&=& \frac{G_F}{\sqrt{2}} \Biggl\{ V_{ub} V_{us}^* a_2 -V_{cb}
V_{cs}^* a_2 \Delta i_5 \Bigl( \frac{m_{\eta^{(')}}^2}{m_c^2} \Bigr)
\nonumber \\
&-&V_{tb} V_{ts}^* \Biggl[ 2(a_3 -a_5)
+\frac{1}{2}(a_9-a_7) + \Biggl( a_3 -a_5 -\frac{1}{2}(a_9-a_7)+a_4 -
\frac{1}{2} a_{10} \nonumber \\
&& 
-(a_6-\frac{1}{2}a_8)\frac{m_{\eta^{(')}}^2}{m_s (m_b + m_s)} \Biggr) 
\frac{f_{\eta^{(')}}^s}{f_{\eta^{(')}}^u} +(a_6-\frac{1}{2}a_8)\frac{ 
m_{\eta^{(')}}^2}{m_s (m_b + m_s)} \nonumber \\
&+& (a_4 -\frac{1}{2} a_{10})
\frac{F_1^{B\rightarrow \eta^{(')}} (m_{K^*}^2)}{A_0^{B\rightarrow
K^*} (m_{\eta^{(')}}^2)} \frac{f_{K^*}}{f_{\eta^{(')}}^u} \Biggr]
\Biggr\} \langle \overline{K^{0*}}
|\overline{s} b_- |\overline{B^0}
\rangle \langle \eta^{(')} |\overline{u} u_-|0\rangle.
\end{eqnarray}

It is instructive to compare the matrix elements derived above 
with the corresponding 
expressions in \cite{AG97}, obtained by estimating the intrinsic charm
quark content in the $\eta$, $\eta^{\prime}$ mesons.
Concentrating on the decays $B^\pm \to
(\eta^\prime,\eta)(K^\pm,K^{*\pm})$, 
which were the ones worked out in \cite{AG97}, and
substituting
 \begin{eqnarray}
  \label{corres}
  - \Delta i_5 (m_{\eta^{\prime}}^2/m_c^2) f_{\eta^{\prime}}^u
  &\rightarrow
  &  f_{\eta^{\prime}}^{c}, \nonumber \\
  - \Delta i_5 (m_{\eta}^2/m_c^2) f_{\eta}^u &\rightarrow
  &  f_{\eta}^{c},
\end{eqnarray}
we get (apart from the small electroweak penguin  contributions
neglected  in \cite{AG97} but included above) exactly 
the same expressions for the decay amplitudes as in \cite{AG97}.
Therefore, we have a simple relation between the decay constants
$f_{\eta^{\prime}}^{c}$, $f_{\eta}^{c}$, introduced in the intrinsic
charm content method, and the form factor
$\Delta i_5$ entering via the operator in Eq.~(\ref{newop}).
The idea of intrinsic charm quark
content of $\eta^\prime$ and $\eta$ and the contribution of the  operator in
Eq.~(\ref{newop}) are  related since this operator
comes from the charm quark loop.
 Using the best-fit values of the
$(\eta,\eta^\prime)$-mixing parameters from \cite{FK97}, yielding
$\theta_8 =-22.2^\circ,
\theta_0 =-9.1^\circ, f_8=168 ~\mbox{MeV}, f_0 =157 ~\mbox{MeV}$,
which in turn yields
$f_{\eta^\prime}^u =63.6$ MeV and $f_\eta^u =77.8$ MeV,
the relations in (\ref{corres}) give $f_{\eta^{\prime}}^{c} \sim 
-3.1$ MeV ($-2.3$
MeV) and $f_{\eta}^{c} \sim -1.2$ MeV ($-0.9$ MeV), with $m_c$ having
the value $ 1.3$ GeV ($1.5$ GeV).
The QCD-anomaly method gives
results for $f_\eta^c$ which are in good agreement with the ones in
\cite{AG97} (in particular for $m_c=1.5$ GeV) but it yields
typically a factor 2 smaller value for $f_{\eta^\prime}^c$ than the
method based on the $(\eta_c,\eta^\prime,\eta)$-mixing \cite{AG97}.
 Given the uncertainties in the fit parameters and approximations in
both the methods, our estimates presented here are consistent with the
parameters $f_{\eta^{\prime}}^{c}$ and $f_{\eta}^{c}$ 
obtained in \cite{AG97}.

However note that the two approaches
are quite different. The advantage of the present approach is that
the operator in Eq.~(\ref{newop}) gives an unambiguous sign relative
to the other contributions while the method of determining the
intrinsic charm quark content via radiative decays $J/\psi \to
(\eta_c,\eta^\prime,\eta)  \gamma$ can only give the absolute
magnitude. In our approach the relative signs of the contributions from
$O_1^c$ and $O_2^c$ to the other contributions are determined;
we obtain
the negative-$f_{\eta^{\prime}}^{c}$ (and 
$f_{\eta}^{c}$) 
solution of the two possible ones which were not resolved in \cite{AG97}.

\vspace{0.3cm}
{\bf 4. Numerical results}

\vspace{0.3cm}
For the numerical analysis, we take the values of the parameters used
in \cite{AG97}.  The matrix elements listed in $\mathbf
(i)$,...,$\mathbf (iv)$ depend on the effective coefficients
$a_1,...,a_{10}$, quark masses, various form factors, coupling
constants and the CKM parameters. 
In turn,  the  coefficients $a_i$ and the quark masses depend on
the renormalization scale $\mu$ and
the QCD scale parameter $\Lambda_{\overline{\mbox{MS}}}$.
We have fixed $\Lambda_{\overline{\mbox{MS}}}$ using $\alpha_s
(m_Z)=0.118$, which is
the central value of the present world average
 $\alpha_s (m_Z)=0.118 \pm 0.003$ \cite{Schmelling96}. The
scale $\mu$ is varied between $\mu =m_b$ and $\mu =m_b/2$, but due to
the inclusion of the NLL expressions, the dependence of the decay
rates on $\mu$ is small and hence  not pursued any further.
To be definite, we use $\mu=2.5$ GeV in the following..
The CKM matrix will be
expressed in terms of the Wolfenstein parameters \cite{Wolfenstein83},
$A$, $\lambda$, $\rho$ and $\eta$. Since the first two are
well-determined
with $A= 0.81 \pm 0.06, ~\lambda=\sin \theta_C=0.2205 \pm 0.0018$, we
fix them to their central values. The other two are correlated and are
found to lie (at 95\% C.L.) in the range $0.25 \leq \eta \leq 0.52$
and $-0.25 \leq \rho \leq 0.35$ from the CKM unitarity fits
\cite{AL96}. However, a good part of the negative-$\rho$ region is now 
disfavoured \cite{paganini97}
by the lower bound on the mass mixing ratio $\Delta M_s/\Delta M_d$
\cite{LEPxs}. Likewise, the ratio $R_1=0.65 \pm 0.40$ measured
recently by the CLEO collaboration \cite{CLEOkpi}, with $R_1 \equiv
{\cal B} (B^0(\bar{B^0}) \to \pi^\pm K^\mp)/{\cal B}(B^\pm \to \pi^\pm
K^0)$, disfavours the region $\rho < 0$ \cite{AG97}.
 Hence, we shall not entertain here the
negative-$\rho$ values and take three representative points
in the allowed $(\rho,\eta)$ contour with $\rho \geq 0$. These  
correspond to the three values of the CKM matrix element ratio:
 $|V_{ub}/V_{cb}|=
0.08, ~0.11, ~\mbox{and} ~0.05$, reflecting the present
central value of this quantity and the upper and lower values
resulting from a generous error on it.
 The specific values of $\rho$ and $\eta$ and the legends used in 
drawing the figures  are as follows:
\begin{enumerate}
\item $\rho = 0.05, \eta = 0.36$, yielding $\sqrt{\rho^2 + \eta^2}
=0.36 $ (drawn as a solid curve)
\item $\rho = 0.30, \eta = 0.42$, yielding
$\sqrt{\rho^2 + \eta^2} =0.51 $ (drawn as a dashed curve)
\item $\rho = 0, \eta = 0.22$, yielding
$\sqrt{\rho^2 + \eta^2} =0.22 $ (drawn as a dashed-dotted curve).
\end{enumerate}
All other curves in Figs.~$2$ - $4$, through which other parametric 
dependences 
are shown, are based on using the values $\rho=0.05,~\eta=0.36$. 
 The decay constants
$f_{\eta^{\prime}}^u$, $f_{\eta^{\prime}}^s$, $f_{\eta}^u$ and
  $f_{\eta}^s$ can be obtained from $f_0$ and $f_8$ by using
$\theta_0$ and $\theta_8$ for the $\eta^{\prime}$-$\eta$ mixing
angles. Since   $q^2 = m_h^2$ is rather
close to the point $q^2 =0$, and a simple pole model is mostly used to
implement the $q^2$ dependence in the form factors, we shall neglect
the $q^2$ dependence and equate $F_{0,1}^{B \to h} (q^2 =m_h^2) =
F_{0,1}^{B \to h} (q^2 =0)$. The values used for
the form factors  are listed in Table \ref{table1}.
\begin{table}[htb]
\begin{center}
\begin{tabular}{| r | r | r | r|}
\hline
 $F_{0,1}^{B \to K}$ & $F_{0,1}^{B \to \eta^\prime}~~~~~~$
& $F_{0,1}^{B \to \eta}~~~~~~$  & $A_0^{B \to K^*}$ \\
 \hline
$0.33$ & $0.33 \left[ \frac{\sin \theta_8}{\sqrt{6}} + \frac{\cos 
\theta_0}{\sqrt{3}} \right] $ &
$0.33 \left[ \frac{\cos \theta_8}{\sqrt{6}} - \frac{\sin
 \theta_0}{\sqrt{3}} \right] $ & 0.28 \\
\hline
\end{tabular}
\end{center}
\caption{Form factors at $q^2=0$.}
\label{table1}
\end{table}

The quark masses enter our analysis in two different ways.
First, they occur in the amplitudes involving penguin loops.
We treat the internal quark masses in these loops
as constituent masses rather than current masses. For them we use the 
following (renormalization scale independent) values:
\begin{equation}
m_b=4.88 ~\mbox{GeV}, ~~m_c=1.5 ~\mbox{GeV}, ~~m_s=0.5 ~\mbox{GeV},
~m_d=m_u=0.2  ~\mbox{GeV}.
\end{equation}
Variation in a reasonable range of these parameters does not change
the numerical results of the branching ratios significantly. The value
of $m_b$ above is fixed to be the current quark mass value
$\overline{m_b}(\mu=m_b/2)=4.88$ GeV, given below.
 Second, the quark masses $m_b$, $m_s$, $m_d$ and $m_u$
also appear through the equations of motion
when working out the (factorized) hadronic matrix elements.
In this case, the quark masses should be interpreted as current masses.
Using $\overline{m_b}(m_b)=4.45 \, $ GeV \cite{GKL96}
and
\begin{equation}
\label{msbarmass}
\overline{m_s}(1 \ \mbox{GeV}) = 150 \ \mbox{MeV} \quad , \quad
\overline{m_d}(1 \ \mbox{GeV}) = 9.3 \ \mbox{MeV} \quad , \quad
\overline{m_u}(1 \ \mbox{GeV}) = 5.1 \ \mbox{MeV} \quad ,
\end{equation}
from \cite{GaL85}, the corresponding values at the renormalization
scale $\mu=2.5$ GeV are given in Table 2, together with other
input parameters needed for our analysis.
\begin{table}[htb]
\begin{center}
\begin{tabular}{| r | r | r | r | r | r |}
\hline
 $\overline{m_b}~~~$ & $\overline{m_s}~~~$  &
$\overline{m_d}~~~$ & $\overline{m_u}~~~$ & $\alpha_s(m_Z)$
& $\tau_B~~~$  \\
 \hline
4.88 GeV  & 122 MeV & 7.6 MeV & 4.2 MeV & 0.118 & 1.60 ps  \\
\hline
\end{tabular}
\end{center}
\caption{Quark masses and other input parameters. The running masses
are given at the renormalization scale $\mu=2.5$ GeV in the 
$\overline{\mbox{MS}}$ scheme.}
\label{table2}
\end{table}
\begin{table}[htb]
\begin{center}
\begin{tabular}{|c|c|c|c|}  \hline
Processes & BR ($\xi=0$)& BR ($\xi=0.45)$ & Experiment
\\ \hline
$B^{\pm} \rightarrow \eta^{\prime} K^{\pm}$&
$(2.7 - 3.6) \times 10^{-5}$ &$(2.0 - 2.6) \times 10^{-5}$ & 
$(6.5^{+1.5}_{-1.4} \pm 0.9)\times 10^{-5}$ \\ \hline
$B^{\pm} \rightarrow \eta^{\prime}K^{*\pm}$ &
$(2.6 - 5.2) \times 10^{-7}$& $(2.0 - 3.9) \times 10^{-7}$& $<
1.3\times 10^{-4}$ \\ \hline
$B^{\pm} \rightarrow \eta K^{\pm}$&
$(1.5 - 2.2) \times 10^{-6}$ & $(0.8 - 1.7) \times 10^{-6}$ &
$<1.4\times  10^{-5}$ \\ \hline $B^{\pm} \rightarrow \eta K^{*\pm}$ & 
$(1.5 - 3.0) \times 10^{-6}$&$(1.6 - 3.1) \times 10^{-6}$&$<3.0\times
10^{-5}$ \\ \hline
$B^{0} \rightarrow \eta^{\prime} K^{0}$&
$(3.0 - 3.7) \times 10^{-5}$ &$ (2.0 - 2.6) \times 10^{-5}$ &
$(4.7^{+2.7}_{-2.0} \pm 0.9)\times 10^{-5}$ \\ \hline
$B^{0} \rightarrow \eta^{\prime}K^{*0}$ &
$(2.0 - 6.5) \times 10^{-7}$& $(0.8 - 1.0) \times 10^{-7}$& $<
3.9\times 10^{-5}$ \\ \hline
$B^{0} \rightarrow \eta K^{0}$&
$(1.2 - 1.7) \times 10^{-6}$ & $(0.8 - 1.1) \times 10^{-6}$ & 
$< 3.3 \times 10^{-5}$
 \\ \hline $B^{0} \rightarrow \eta K^{*0}$ &
$(2.4 - 3.4) \times 10^{-6}$&$(1.9 - 2.8) \times 10^{-6}$&$<3.0\times 
10^{-5}$ \\ \hline 
\end{tabular}
\end{center}
\caption{Numerical estimates of the branching ratios for the decays
$B^\pm \to (\eta^\prime,\eta)(K^\pm,K^{*\pm})$ and $B^0 \to 
(\eta^\prime,\eta)(K^0,K^{*0})$, obtained by varying the  CKM 
parameters in the ranges indicated in the text and the $s$-quark mass
in the range $100 ~\mbox{MeV} \leq \overline{m_s}( 2.5 ~\mbox{GeV})
\leq 122 ~\mbox{MeV}$. The first column corresponds to using the value
of the factorization-model parameter $\protect\xi=0$ and the second to
$\protect\xi=0.45$. The third column shows the present measurement and
the $90\%$ C.L. upper limits reported by the CLEO collaboration
\protect\cite{CLEO1,CLEO3}.} 
\label{table3}
\end{table}

   The branching ratios $BR(B^{\pm} \rightarrow \eta^{\prime} K^{\pm})$
and $BR(B^{0} \rightarrow \eta^{\prime} K^{0})$
are plotted against the\\ parameter $\xi$ in Figs.~2 and
3, respectively. 
Although not indicated by the notation, the branching ratios for the
neutral $B$-meson decays are always understood to be averages with the
corresponding charged conjugated decays in the following tables and figures. 
We would like to make the following observations
concerning the sensitivity of the branching ratios on the various input 
parameters.
\begin{itemize} 
\item CKM-parametric dependence:  The branching ratio  
$BR(B^{\pm} \rightarrow \eta^{\prime} K^{\pm})$
 shows a mild dependence (of order $10\%$) on the CKM parameters 
whereas  $BR(B^{0} \rightarrow \eta^{\prime} K^{0})$ is practically 
independent of them.
\item $s$-quark mass dependence:
 The dependence of the branching ratios on the $s$-quark mass is
quite marked. To illustrate
this we show the two branching ratios calculated using
$\overline{m_s}( 2.5 ~\mbox{GeV}) = 100 $ MeV through the dotted
curves, to be compared with the corresponding solid curves which are
drawn for the same values of the CKM parameters, namely $\rho=0.05$
and $\eta=0.36$, but $\overline{m_s}( 2.5 ~\mbox{GeV}) = 122 $
MeV. The resulting increase in the branching ratios in lowering the
value from $\overline{m_s}=122$ MeV to $\overline{m_s}=100$ 
MeV amounts to about $20\%$ (and $65\%$ for $\overline{m_s}=80$ MeV). 
While in literature one comes across even
smaller values of $\overline{m_s}( 2.5 ~\mbox{GeV})$, we subscribe to the 
view 
that the value $\overline{m_s}( 2.5 ~\mbox{GeV})=100$ MeV is a realistic 
lower limit on this
quantity. However, if the present high values of the branching ratios
for $B^{\pm} \rightarrow \eta^{\prime} K^{\pm}$ and 
$B^{0} \rightarrow \eta^{\prime} K^{0}$ continue to persist, one might
have to consider smaller values of $\overline{m_s}$.
We also expect  progress in calculating quark masses on the
lattice, sharpening the theoretical estimates presented here.
\item Dependence on $f_{\eta^\prime}^c$:
Restricting to the negative-$f_{\eta^\prime}^c$ 
solution, determined by the QCD-anomaly method, we plot the branching 
ratios with $f_{\eta^\prime}^c=-5.8$ MeV, as obtained in
\cite{AG97}. The other parameters are: $\overline{m_s}( 2.5
~\mbox{GeV}) = 100 $ MeV, $\rho=0.05$ 
and $\eta$=0.35. The results are shown through the long-short dashed
curves in Figs.~2 and 3. Comparing these curves with the corresponding
dotted curves, we see that the resulting branching ratios in the two
approaches are very similar.
\item Dependence on $\xi$: This amounts to between 20\% and 35\%
depending on the other parameters if one varies $\xi$ in the
range $0 \leq \xi \leq 0.5$. In all cases, the branching ratios are
larger for $\xi=0$. 
\end{itemize}
Taking into account the parametric dependences just discussed,
we note that the 
theoretical branching ratios $BR(B^{\pm} \rightarrow \eta^{\prime}
K^{\pm})$ and $BR(B^{0} \rightarrow \eta^{\prime} K^{0})$ are
uncertain by a factor 2. 
Determining the value of $\xi$ from other $B \to h_1 h_2$ decays in
the future, this uncertainty can be reduced considerably (see Table 
\ref{table3}).

 The ratio of the branching ratios
$BR(B^{\pm} \rightarrow \eta^{\prime} K^{\pm})/BR(B^{0} \rightarrow 
\eta^{\prime} K^{0})$  is a useful quantity, as
it is practically independent of the form factors and most input   
parameters. This is shown in Fig.~4. We see that the residual
uncertainty is due to the CKM-parameter dependence of this ratio,
which is estimated as about 10\%. We get (for $0 \leq \xi \leq
0.5$) 
\begin{equation}
\label{ratior1}
\frac{BR(B^{\pm} \rightarrow \eta^{\prime} K^{\pm})}{BR(B^{0}
\rightarrow \eta^{\prime} K^{0})} = 0.9 - 1.02 ~.
\end{equation}  
The present experimental value of this ratio as calculated by adding
the experimental errors in the numerator and denominator in quadrature
is  $1.38 \pm 0.86$. Given the large experimental error, 
it is difficult to draw any quantitative
conclusions except that the theoretical ratio in Eq.~(\ref{ratior1})
is in agreement  with data. However, we do expect that the
experimental value of this ratio will asymptote to unity.

The branching ratio $BR(B^{\pm} \rightarrow \eta^{\prime} K^{\pm})$ is
found to be somewhat lower than the present experimental
number reported by CLEO. As shown in Fig.~2, we estimate 
$BR(B^{\pm} \rightarrow \eta^{\prime} K^{\pm})=(2 
-4) \times 10^{-5}$ in our framework compared to the experimental
measurement of the same $(6.5 \pm 1.75) \times 10^{-5}$. The
calculations presented here are in better agreement with the 
experimental measurement of $BR(B^{0}\rightarrow \eta^{\prime}
K^{0})$, making it to within $1 \sigma$. Of course, theoretical rates
can be augmented by increasing the values of the input form factors
given in Table \ref{table1}. There is certainly some room for such
enhancement, but in view of the emerging theoretical    
consensus on the estimates of the form factors and the fact that the 
branching ratios in a number of $B \to h_1 h_2$ decays do not require
such enhancement \cite{AG97}, this can only be modest. Hence, we anticipate 
that the
experimental numbers for $BR(B^{\pm} \rightarrow \eta^{\prime} K^{\pm})$
and $BR(B^{0}\rightarrow \eta^{\prime} K^{0})$
will decrease so as to be more in line with the rest of the
CKM-allowed QCD-penguin-dominated two-body $B$ decays and with our
estimates!
 
  We present in Table \ref{table3} numerical estimates for all the
eight branching ratios $BR(B^{\pm} \rightarrow (\eta^{\prime},\eta)( 
K^{\pm},K^{*\pm}))$ and
$BR(B^{0} \rightarrow (\eta^{\prime},\eta)( K^{0},K^{*0}))$ calculated
in the QCD-anomaly method. The ranges shown take into account
the $m_s$- and CKM-parametric dependence, discussed earlier.
The entries in column 2 and 3 are based on the choice $\xi=0$, 
corresponding to using $N_c=\infty$ in the effective coefficients, and
$\xi=0.45$, which corresponds to the phenomenological value estimated 
in the decays $B \to (D,D^*)(\pi,\rho)$ \cite{NS97}, respectively

A number of comments are in order on the entries in Table
\ref{table3}.  First, as shown in this table 
and Fig.~4, theoretical estimates of the branching ratios for
 $B^{\pm}\rightarrow \eta^{\prime} K^{\pm}$ and $B^{0} \rightarrow
\eta^{\prime} K^{0}$ are almost equal and they are also the largest
for the eight decays shown. So, it is no coincidence that these are
exactly the ones measured so far. In particular, the branching ratios
for the decays $B^{\pm}\rightarrow \eta^{\prime} K^{*\pm}$ and $B^{0}
\rightarrow \eta^{\prime} K^{*0}$ are found to be the smallest in this
group, and we predict 
\begin{eqnarray}
\label{kkstar}
\frac{BR(B^{\pm}\rightarrow \eta^{\prime}
K^{*\pm})}{BR(B^{\pm}\rightarrow  
\eta^{\prime} K^{\pm})} &\leq & 0.02 ~,\\
\frac{BR(B^{0}\rightarrow \eta^{\prime} 
K^{*0})}{BR(B^{0}\rightarrow    
\eta^{\prime} K^{0})} & \leq & 0.03~.
\end{eqnarray}
Likewise, the branching ratios for the decay modes $B^{\pm}\rightarrow 
\eta K^{\pm}$ and $B^{0} \rightarrow \eta K^{0}$ are smaller compared
to  their $\eta^\prime$-counterparts by at least an order of magnitude.
We estimate  $ BR(B^{\pm}\rightarrow \eta K^\pm) =(1 - 2) \times
10^{-6}$ and a similar value for the neutral $B$ decay mode.
On the other
hand, the branching ratios for the decay modes $B^\pm \to \eta 
(K^\pm,K^{*\pm})$  
and $B^0 \to \eta (K^0,K^{*0})$ are all comparable to each other and
are predicted to be somewhere in the range $(1 - 3) \times 10^{-6}$.
The reason behind this pattern can be seen in the various constructive
and destructive interferences involving the eight amplitudes listed
earlier. This is in qualitative agreement with the anticipation in
\cite{Lipkin}. 

\vspace{0.3cm}
{\bf 5. Concluding Remarks}

\vspace{0.3cm}
We have provided an independent estimate of
the process $b \to s(c\bar{c}) \to s (gg) \to s (\eta^\prime,\eta)$,
using QCD anomaly. The resulting branching ratios in $B \to 
(\eta^\prime,\eta)(K,K^*)$ reported here are  
close to the ones  obtained in the $(\eta_c,\eta^\prime,\eta)$-mixing 
approach  \cite{AG97}. The present method also
removes the intrinsic sign ambiguity inherent in ref.~\cite{AG97},
thereby reducing one source of calculational uncertainty.
Theoretical predictivity is, however, still limited due to 
the remaining input parameters of 
this framework and we estimate it to be a factor 2.
Despite this, very clear patterns emerge among the various decay modes
considered here, which  are drastically
different from the ones which follow in other scenarios. Hence,
ongoing and future experiments will be able to test the 
predictions of the present approach as well as of the
competing ones, such as models based on the dominance 
of the intrinsic charm contributions in $\eta^\prime$, as suggested in 
\cite{HZ97,SZ97}, 
or models in which dominant role is attributed 
to the soft-gluon-fusion process to form an $\eta$ or $\eta^\prime$ 
\cite{Mohd97,Kim97}. In contrast to our approach, 
these models typically predict  similar 
branching ratios (within a factor 2) involving the modes $B \to 
\eta^\prime K$ and $B \to 
\eta^\prime K^*$, in both the charged and neutral $B$ decays. Data
already rules out models predicting $BR(B \to \eta^\prime K^*) > BR(B
\to  \eta^\prime K)$, and is tantalizingly close to
testing also the soft-gluon fusion models which
predict $BR(B \to \eta^\prime K^*) \simeq 0.5 BR(B \to 
\eta^\prime K)$ \cite{Mohd97,Kim97}. We note that a large
intrinsic-charm  component in $\eta^\prime$ is not substantiated by
independent analysis  of the $\eta^\prime \gamma$ form factor
\cite{FK97}. Soft-gluon-fusion models are not theoretically motivated
as they show extreme (quartic) sensitivity of the decay widths to the
gluon mass (an ill-defined quantity) -- reflecting 
that the method being employed in these models is neither infrared stable nor
predictive. 
For a definite test of the theoretical framework presented here and in
\cite{AG97}, more precise data are required and one has to reduce the 
uncertainty on the input parameters, in particular the $s$-quark mass.
 However, if in forthcoming experiments, the branching ratios  
presented in Table \ref{table3} for the eight decay modes are found to
be significantly larger (in particular in the modes $B^\pm \to \eta^\prime 
K^{*\pm}$ and $B^0 \to \eta^\prime K^{*0}$), then it will most probably
be an indication of significant non-factorizing contributions, which
may feed into the decays $B \to (\eta^\prime,\eta)(K,K^*)$ 
dominantly through the matrix elements of the dipole operator $O_g$.

 In conclusion, we reiterate the
two intrinsic assumptions of our approach: (i) absence of final state
interactions, and (ii) absence of non-perturbative contributions in
penguin diagrams, which may modify both the magnitudes and phases of
the effective coefficients calculated in the factorization framework
presented here. 
The first should be a good approximation for the decays being
considered. For the second, we note that non-perturbative
contributions are not specific to the decays $B \to
(\eta^\prime,\eta)(K,K^*)$  but are endemic to all such
decays where penguins play a dominant role \cite{martinellicharm}.
Furthermore, we admit that the factorization framework used here
and elsewhere is vulnerable and it is conceivable
that non-perturbative non-factorizing contributions do play
a significant role in non-leptonic two-body $B$-decays being
discussed here. This remains to be tested experimentally or proven
on firm theoretical grounds in a well-defined framework, such as
lattice QCD. However, it is fair to say that there exists at least
a {\it prima facie} case in some of the related decays, such as 
$B \to K \pi$ and $B \to \pi \pi$, 
which support the contention that the neglected  
non-perturbative effects are not overwhelming and 
the measured decay rates can be explained without invoking them
\cite{AG97}. Of course, this point of view may have to be revised
with more precise data.

\section*{Acknowledgment}
JC and PK are grateful to DESY for hospitality, where
this work was initiated. AA would like to thank Professor H.S.~Song
for the hospitality
at Seoul National University. This work has been partially
supported by the German-Korean scientific exchange programme 
DFG-446-KOR-113/72/0 and by Schweizerischer Nationalfonds. 
JC and PK were supported in part by the Ministry
of Education grants BSRI 97-2408 and BSRI 97-2418, respectively, and
the Korea Science and Engineering Foundation (KOSEF) through the SRC
program of SNU-CTP, and the Distinguished Scholar Exchange Program of
Korea Research Foundation. 
PK is also supported in part by KOSEF, Contract 971-0201-002-2.
We would like to thank Thorsten Feldmann,
Gustav Kramer, Heiri Leutwyler, Peter Minkowski, Hubert Simma and Jim Smith for
discussions.

\begin{figure}[p]
\centerline{
\epsfig{file=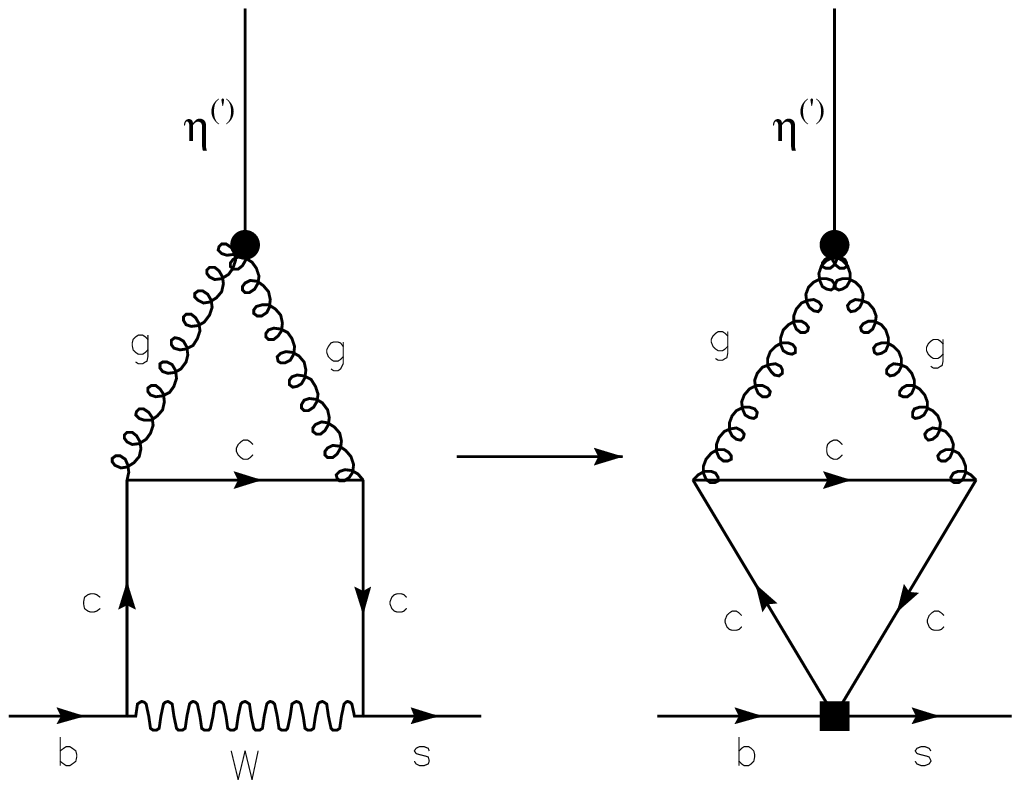,
height=3.0in,angle=0,clip=}
}
\caption[]{
Feynman diagram contributing to the processes $b \to s
(c\bar{c}) \to s (gg) \to s \eta^{(')}$ in the full and effective theory.
 The lower vertex in the diagram on the right is
calculated with the insertion of the operators $O_1^c$ and $O_2^c$ in
the effective Hamiltonian approach; the upper vertex 
in both the full and effective theory is determined
by the QCD triangle anomaly. 
\label{fig1}}
\end{figure}

\begin{figure}[p]
\centerline{
\epsfig{file=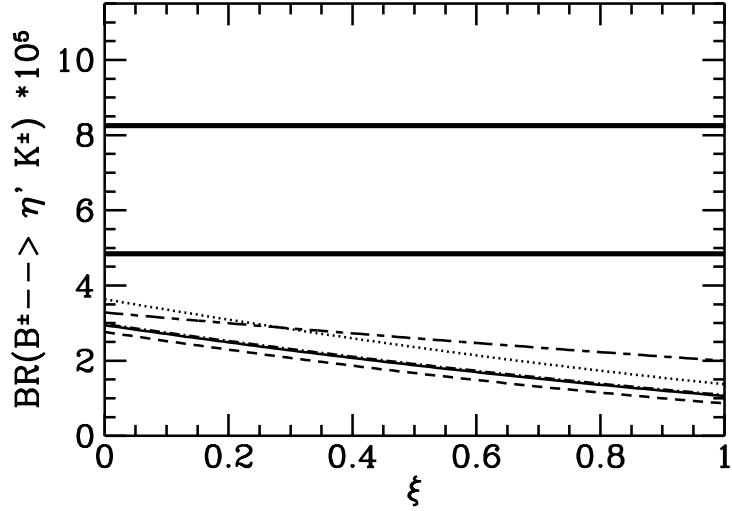,
height=3.0in,angle=0,clip=}
}
\caption[]{
The branching ratio $BR(B^\pm \to \eta^\prime K^\pm)$
plotted against the parameter $\xi$. The lower three curves correspond
to the value $\overline{m_s}( 2.5 ~\mbox{GeV}) = 122 $ MeV and
the three choices of the CKM parameters:
 $\rho = 0.05, \eta = 0.36$ (solid curve);
$\rho = 0.30, \eta = 0.42$ (dashed curve);
$\rho = 0, \eta = 0.22$ (dashed-dotted curve).
The upper two curves correspond to the value $\overline{m_s}( 2.5 
~\mbox{GeV}) = 100 $ MeV, $\rho = 0.05, \eta = 0.36$ and
$f_{\eta^\prime}^c  =-2.3$
MeV from the QCD-anomaly method (dotted curve) and $f_{\eta^\prime}^c
=-5.8$ MeV from \protect\cite{AG97} (long-short dashed curve). The
horizontal thick solid lines represent the present CLEO measurements
(with $\pm 1 \sigma$ errors). 
\label{fig2}}
\end{figure}

\begin{figure}[p]
\centerline{
\epsfig{file=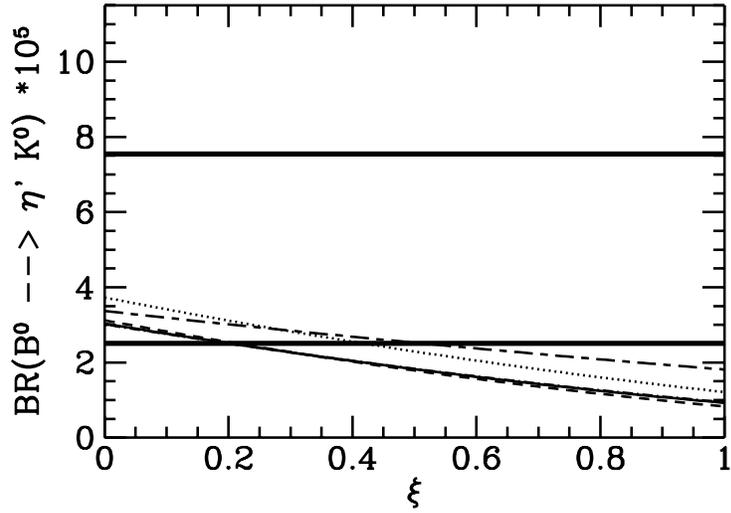,
height=3.0in,angle=0,clip=}
}
\caption[]{
The branching ratio $BR(B^0 \to \eta^\prime K^0)$ plotted
against the parameter $\xi$. The legends are the same as in Fig.~2.
The horizontal thick solid lines represent the present CLEO
measurements (with $\pm 1 \sigma$ errors).
\label{fig3}}
\end{figure}

\begin{figure}[p]
\centerline{
\epsfig{file=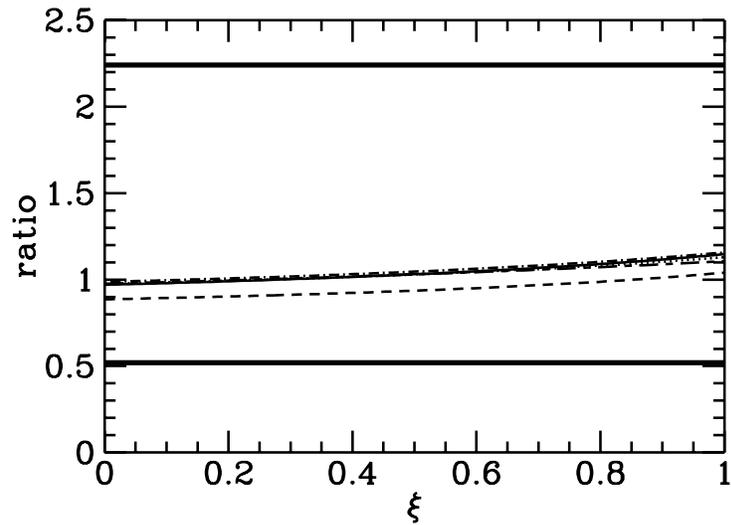,
height=3.0in,angle=0,clip=}
}
\caption[]{
Ratio of the branching ratios $BR(B^\pm \to \eta^\prime
K^\pm)$/ $BR(B^0 \to \eta^\prime K^0)$ plotted against the parameter
$\xi$. The legends are the same as in Fig.~2.  The horizontal thick
solid lines represent the present CLEO measurements (with $\pm 1
\sigma$ errors estimated as stated in the text).
\label{fig4}}
\end{figure}
\end{document}